\documentclass[journal]{IEEEtran}

\ifCLASSINFOpdf
\else
   \usepackage[dvips]{graphicx}
\fi
\usepackage{url}

\usepackage{graphicx}
\usepackage{soul,color}
\usepackage{amsmath}
\usepackage{multirow}
\usepackage{hyperref}

\usepackage{subcaption}

\def\etal{\emph{et al}.\ }

\begin{document}

\title{Reverberation-based Features for Sound Event Localization and Detection with Distance Estimation}

%\author{Davide Berghi, \IEEEmembership{Member, IEEE} and Philip J. B. Jackson, \IEEEmembership{Member, IEEE}
\author{Davide Berghi \IEEEmembership{Member, IEEE} and Philip J. B. Jackson, \IEEEmembership{Member, IEEE}
\thanks{%This paragraph of the first footnote will contain the date on which you submitted your paper for review. It will also contain support information, including sponsor and financial support acknowledgment. 
Research supported by EPSRC and BBC Prosperity Partnership AI4ME: Future Personalised Object-Based Media Experiences Delivered at Scale Anywhere EP/V038087. For the purpose of open access, the author has applied a Creative Commons Attribution (CC BY) licence to any Author Accepted Manuscript version arising. Data supporting this study are available from \url{https://zenodo.org/records/7880637} and from \url{https://zenodo.org/records/10932241}.}
\thanks{D. Berghi and P. J. B. Jackson are with the Centre for Vision, Speech, and Signal Processing, University of Surrey, Guildford, U.K. \\(e-mail:\ p.jackson@surrey.ac.uk).}}

%\markboth{Journal of \LaTeX\ Class Files, Vol. 14, No. 8, August 2015}
%{Berghi \MakeLowercase{\textit{et al.}}: Bare Demo of IEEEtran.cls for IEEE Journals}
\maketitle

\begin{abstract}
%Sound event localization and detection (SELD) consists of predicting the classes of active sound events over time while estimating their positions. Traditionally, the localization subtask in SELD has been addressed as a direction of arrival estimation problem. %, focusing on predicting the azimuth and elevation components of the incoming sound event's trajectory. However, this approach does not account for the distance of the audio sources. %which plays an important role in many real applications. To address this limitation, recent advancements have extended. Only recently, SELD was extended to include distance estimation, enabling the prediction of sound event positions in 3D space (3D SELD). However, existing works on 3D SELD have yet to propose an input feature format specifically designed for distance estimation. We hypothesize that valuable information about sound event distances is encoded in the reverberation of the input signals. This paper introduces two novel feature formats derived from audio reverberation. The first utilizes direct-to-reverberant ratio (DRR) information, while the second leverages signal autocorrelation functions to provide the model with insights into early reflections. After pre-training on synthetic data, the model shows improvements in relative distance error (RDE) and overall SELD score with both methods. Specifically, with autocorrelation-based features, the RDE was reduced by over 3 percentage points on the STARSS23 dataset.

Sound event localization and detection (SELD) involves predicting active sound event classes over time while estimating their positions. The localization subtask in SELD is usually treated as a direction of arrival estimation problem, ignoring source distance. Only recently, SELD was extended to 3D by incorporating distance estimation, enabling the prediction of sound event positions in 3D space (3D SELD). However, existing methods lack input features specifically designed for distance estimation. We address this gap by introducing two novel reverberation-based feature formats: one using the direct-to-reverberant ratio (DRR) and another leveraging signal autocorrelation to capture early reflections.
%This paper introduces two novel feature formats for 3D SELD based on reverberation: one using direct-to-reverberant ratio (DRR) and another leveraging signal autocorrelation to provide the model with insights into early reflections.
%Pre-training on synthetic data improves relative distance error (RDE) and overall SELD score, with autocorrelation-based features reducing RDE by over 3 percentage points on the STARSS23 dataset.
We extensively evaluate and benchmark these features on the STARSS23 dataset, combining them with established SELD features for sound event detection (SED) and direction-of-arrival estimation (DOAE), and testing across different network architectures. Our proposed features, applicable to both FOA and MIC formats, achieve state-of-the-art distance estimation, enhancing overall 3D SELD performance. %Pre-training on synthetic data followed by fine-tuning on real recordings further improves robustness, reducing relative distance error (RDE) by over 3 percentage points and delivering balanced 3D SELD performance

\end{abstract}

%\vspace{-1mm}

\IEEEpeerreviewmaketitle

\section{Introduction}
\label{sec:intro}

\IEEEPARstart{S}{ound} event localization and detection (SELD) \cite{Adavanne:2019:SELDnet} integrates two subtasks: sound event detection (SED) and sound source localization (SSL). Thus, it involves identifying active sound events at any given time frame while estimating their spatial positions. SELD systems are important in many practical applications, e.g., human-robot interaction, security, accessibility, and safety.
SELD gained significant attention following its inclusion as a task in the Detection and Classification of Acoustic Scenes and Events (DCASE) Challenge.
Recent advancements in SELD research have tackled increasingly complex challenges, such as detecting moving events \cite{politis:2020:DCASE}, ignoring external interfering sounds \cite{politis:2021:DCASE}, distinguishing simultaneous same-class events originating from different directions \cite{cao:2020:EIN,cao:2021:EINv2,Shimada:2022:multiACCDOA}, and leveraging the visual modality to tackle SELD as a multimodal task \cite{Shimada2023STARSS23AA,Berghi:2024:ICASSP24,Jiang:2024:AVseld,Roman:2024:ehnancedAVseld}.

%The localization aspect of SELD is traditionally framed as a direction of arrival estimation (DOAE) problem, predicting the azimuth and elevation of sound events. However, this overlooks source distance, a crucial factor in many applications. The DCASE 2024 challenge addressed this by introducing distance estimation (3D SELD) \cite{Diaz-Guerra:2024:seldBaseline24}.
The localization aspect of SELD is typically treated as a direction-of-arrival (DOA) estimation task, predicting sound event azimuth and elevation. However, this ignores source distance, which is critical in many applications. The DCASE 2024 challenge addressed this by introducing distance estimation (3D SELD) \cite{Diaz-Guerra:2024:seldBaseline24}.
Krause \etal \cite{Krause:2024:seldDistance} proposed two methods to support distance estimation in 3D SELD. The first extends the multi-activity-coupled Cartesian DOA (multi-ACCDOA) vectors \cite{Shimada:2022:multiACCDOA} to include distance estimation, predicting for each event class \textit{c}, track \textit{n}, and time frame \textit{t}, a 3D DOA vector -- (\textit{x, y, z}) coordinates on the unit sphere -- along with a distance value $D_{nct} \in \langle0,\infty)$. This representation, referred to as the multi-activity-coupled Cartesian Distance and DOA (multi-ACCDDOA) method, incorporates distance estimation into the original framework. The second method presented in \cite{Krause:2024:seldDistance} adds a separate output branch for distance estimation.
Hong \etal \cite{hong:2024:MVAnet} instead predict full 3D positions of sound events (\textit{x, y, z}), combining DOA and distance into a single representation. However, this requires an additional output branch to handle the SED subtask. Dong \etal \cite{dong:2025:anExperimStudy} adopt a multi-branch ResNet-Conformer architecture for 3D SELD. Abolfazli \etal \cite{Abolfazli:2024:3dseld_withresnet} report slight distance improvements, replacing the CNN backbone of the DCASE baseline with a pre-trained ResNet backbone, but with degraded performance on other SELD metrics.
Sato \etal \cite{Sato:2025:physicsInformedDist} introduce a physics-informed architecture for distance estimation, achieving only marginal gains over the baseline.

The methods presented above primarily address the task from an architectural perspective, sometimes even degrading overall SELD performance. None of these approaches investigates dedicated input features for 3D SELD. Selecting the right input features is crucial in designing a SELD system \cite{Berghi:2023:WASPAA}. Commonly adopted features include log-mel spectrograms for the SED subtask, intensity vectors (IV) \cite{cao:2020:EIN} for DOAE in first-order ambisonics (FOA) audio, and generalized cross-correlation with phase transform (GCC-PHAT) \cite{Knapp:gccphat:1976} or SALSA/SALSA-Lite \cite{Nguyen:2021:SALSA,Nguyen:2021:SALSALiteAF} for microphone array (MIC) format. However, features specifically designed for distance estimation have not been extensively explored within the context of 3D SELD.
Useful distance-related information is encoded in acoustic reverberation \cite{Mershon:1975:revDist,Sheeline:1982:DRR,Griesinger:2009:DRR,Lu:2010:binauralDist,Georganti:2011:distDet,chitreddy:2020:distPerc}. Independently and concurrently with our work, Yeow \etal \cite{Yeow:2025:CDPDdistFeat} introduced Coherence and Direct-Path Dominance (CDPD) features for distance estimation. These features integrate spatial coherence and reverberation cues into conventional SELD features for MIC format (SALSA and SALSA-Lite).
In this paper, we present two novel reverberation-based feature extraction methods. The first uses the direct-to-reverberant ratio (DRR) as an indicator of the energy balance between direct and reverberant sound. The second leverages the autocorrelation function to extract information about early reflections and estimate the initial time delay gap (ITDG), the interval between direct sound and the first major reflection. Unlike CDPD, our features can be extracted from both MIC and FOA formats, and in our experiments, they achieved consistently higher 3D SELD performance. Experiments on the STARSS23 dataset \cite{Shimada2023STARSS23AA} demonstrate that incorporating these features alongside established SELD features (e.g., log-mel spectrograms and intensity vectors) consistently improves distance estimation performance while maintaining, and in some cases even improving, other SELD metrics.

%The contributions introduced in this paper are threefold: (1) we present two approaches to extract reverberation-based input features to tackle the distance dimension of the 3D SELD task; (2) we show through a preliminary study that the autocorrelation-based features relate to distance by convolving an audio clip with room impulse responses (RIRs) captured at varying distances, (3) we demonstrate the effectiveness of the proposed features on real data.

The main contributions of this paper are threefold: (1) we propose two methods for extracting reverberation-based input features to address distance estimation in 3D SELD; (2) we conduct a preliminary study showing that autocorrelation-based features capture distance-related information by analyzing how an audio clip interacts with room impulse responses (RIRs) recorded at different distances; (3) we conduct an extensive evaluation on real-world data, combining the proposed distance features with established features for SED and DOAE, testing across different network architectures, and benchmarking against CDPD. Our approach achieves state-of-the-art performance in 3D SELD.
The code to extract our features is available at
\href{https://github.com/dberghi/SELD-distance-features}{\texttt{github.com/dberghi/SELD-distance-features}}

%The remaining of this paper is organized as follows...

%\vspace{-3mm}
\section{Proposed Reverberation-based Features}
\label{sec:features}

%While log-mel spectrograms and intensity vectors (IVs) are effective input features for the SED and DOAE subtasks, they are not designed for estimating the sound event distances. 
%Our goal is to provide the network with additional information to improve distance estimation. In this section, we introduce two main input features specifically designed for this task. 
Log-mel spectrograms and intensity vectors (IVs) are effective for SED and DOAE, but they are not suited for distance estimation. To address this, we introduce two input features specifically designed to enhance distance estimation.

%\hl{to disambiguate between the direct sound, the reflected/reverberant sound and any background noise interference. In order to enable this three-way classification, we seek to enrich the audio inputs based on the available microphone signals.}

%\vspace{-3mm}
\subsection{Direct-to-Reverberant Features}

Distance cues can be extracted from the relationship between the direct and reverberant components of the captured audio signals \cite{Mershon:1975:revDist}. %Specifically, the later tail of the reverberant signal (late reverberation) carries information about the sound's apparent distance \hl{check this}.
%We decided to include both direct and reverberant components (``DR'', for compactness) extracted from the Omnidirectional audio channel to the set of input features, in the form of two additional log-mel spectrograms. %For compactness, we refer to these features as per ``DR''.
To estimate the direct sound, $d(t)$, we employed the Weighted Prediction Error (WPE) dereverberation algorithm \cite{Takuya:2012:WPE} applied to the omnidirectional channel W of the first-order ambisonic (FOA) audio format. We adopted the Python implementation of WPE released by Drude \etal \cite{Drude:2018:nara_wpe} (taps=60; delay=5; iterations=5).
The reverberant component, $r(t)$, is then estimated by subtracting the direct signal from the original signal in the temporal domain.
To extract DRR features as 2D inputs to the model and to enable concatenation with the other SELD features (i.e., log-mel spectrograms and IVs), we calculate the DRR as a function of time and frequency, and then mapped it to log-mel space. The proposed DRR input features, $\mathbf{DRR}^{\mathrm{mel}}$, are defined as:
%The time-frequency DRR features are computed as:
\begin{equation}\label{eq:drr}
    \mathbf{DRR}^{\mathrm{mel}}(t,k)=10\cdot\log_{10}\left(\mathbf{P}^{\mathrm{mel}}_{\mathrm{DRR}}(t,k)\right)
\end{equation}
\begin{equation}\label{eq:mel}
    \mathbf{P}^{\mathrm{mel}}_{\mathrm{DRR}}(t,k)= \sum_{f=0}^{F}\mathbf{H}^{\mathrm{mel}}(k,f)\left(\dfrac{\mathbf{P}_\mathrm{D}(t,f)}{\mathbf{P}_\mathrm{R}(t,f)}\right)
\end{equation}
where $\mathbf{H}^{\mathrm{mel}}$ denotes the mel filter bank, which maps the frequency spectrum to the mel scale, with $k$ being the mel bin index. $\mathbf{P}_\mathrm{D}(t,f)$ and $\mathbf{P}_\mathrm{R}(t,f)$ are the power spectral densities (PSDs) of the direct and reverberant components, respectively. To prevent instability or division by zero, the PSD values were clamped to a small positive constant, $\epsilon{=}1e{-}10$. Mathematically, $\mathbf{P}_\mathrm{D}(f,t)$ and $\mathbf{P}_\mathrm{R}(f,t)$ can be 
% defined as: 
% \begin{equation}\label{eq:psd_d}
%     \mathbf{P}_\mathrm{D}(t,f)=\mathrm{max}(|\mathbf{D}(t,f)|^2, \epsilon) 
% \end{equation}
% \begin{equation}\label{eq:psd_r}
%     \mathbf{P}_\mathrm{R}(t,f)=\mathrm{max}(|\mathbf{R}(t,f)|^2, \epsilon)
% \end{equation}
defined as $\mathbf{P}_\mathrm{D}(t,f){=}\mathrm{max}(|\mathbf{D}(t,f)|^2, \epsilon)$ and  
$\mathbf{P}_\mathrm{R}(t,f){=}\mathrm{max}(|\mathbf{R}(t,f)|^2, \epsilon)$, where $\mathbf{D}(t,f)$ and $\mathbf{R}(t,f)$ are the short-term Fourier transforms (STFTs) of the direct and reverberant components, $d(t)$ and $r(t)$, respectively.

In addition to the DRR features described, we explore a variant where $\mathbf{D}(t,f)$ and $\mathbf{R}(t,f)$ are separately converted into log-mel spectrograms and fed into the model. This approach, introduced in our DCASE2024 Task 3 submission \cite{Berghi:2024:DCASE24techRep}, aims to give the network greater flexibility in learning task-relevant information. We refer to these features as D+R features.

\begin{figure}[tb]
\centerline{\includegraphics[width=\columnwidth]{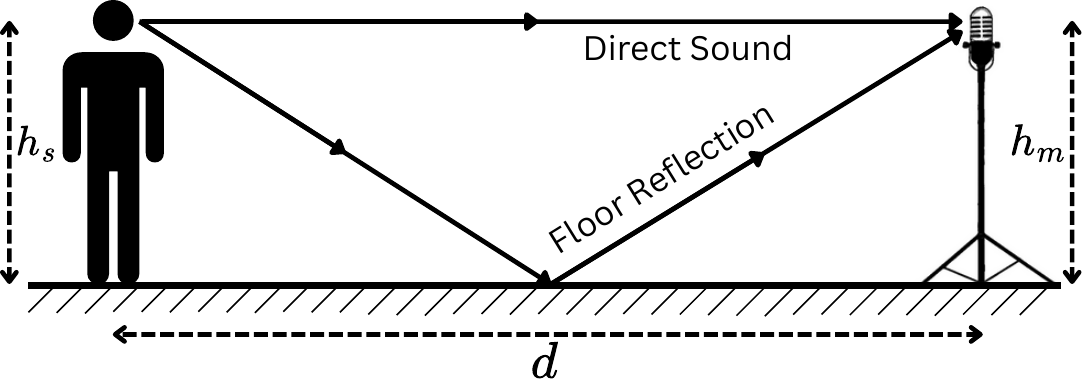}}
\caption{Floor reflection path when source and receiver are at the same height ($h_s{=}h_m$) and separated by distance $d$.}
\label{fig:floorReefl}
\vspace{-3mm}
\end{figure}

\iffalse
\begin{table}[tb]
\caption{Ideal direct sound and first reflection (1stRef) delays, with their corresponding initial time delay gaps (ITDGs), as the source distance increases, assuming the first reflection originates from the floor. We present cases for source heights of 1.5\,m and 0.9\,m, with microphone positioned at 1.5\,m, sound speed of 343\,m/s, and no additional interfering factors.  
%with both the microphone and the source positioned at a fixed height of 1.5\,m, a sound speed of 343\,{m}/s and no additional interfering factors.
}
\begin{center}%\footnotesize
\begin{tabular}{c||c|c|c||c|c|c}
%\hline
\cline{2-7}
& \multicolumn{3}{|c||}{\textbf{Source Height: 1.5m}} & \multicolumn{3}{|c}{\textbf{Source Height: 0.9m}}\\
\hline 
\textbf{Dist}&\textbf{Direct}&\textbf{1stRef}&\textbf{ITDG}&\textbf{Direct}&\textbf{1stRef}&\textbf{ITDG}\\
\hline
1.0\,m & 2.9\,ms & 9.2\,ms & 6.3\,ms & 3.4\,ms & 7.6\,ms & 4.2\,ms\\
1.5\,m & 4.4\,ms & 9.8\,ms & 5.4\,ms & 4.7\,ms & 8.2\,ms & 3.5\,ms\\
2.0\,m & 5.8\,ms & 10.5\,ms & 4.7\,ms & 6.1\,ms & 9.1\,ms & 3.0\,ms\\
2.5\,m & 7.3\,ms & 11.4\,ms & 4.1\,ms & 7.5\,ms & 10.1\,ms & 2.6\,ms\\
3.0\,m & 8.7\,ms & 12.4\,ms & 3.6\,ms & 8.9\,ms & 11.2\,ms & 2.3\,ms\\
%3.5\,m & 10.2\,ms & 13.4\,ms & 3.2\,ms \\
%4.0\,m & 11.7\,ms & 14.6\,ms & 2.9\,ms \\

\hline
\end{tabular}
\label{tab:distances}
\end{center}\vspace{-4mm}
\end{table}
\fi

\subsection{Short-term Power of the Autocorrelation}

%-------------------

%For the second feature, we examined how early reflections contribute to distance estimation, with a particular focus on the ITDG, which plays a key role in distance perception and estimation \cite{Bronkhorst:1999:distPercep,kaplanis2014perception}.
%Although the time delay of early reflections depends on room size, it is reasonable to assume that the earliest reflection originates primarily from the floor \cite{kaplanis2014perception}, as illustrated in Fig.\,\ref{fig:floorReefl}. Based on this assumption, Table\,\ref{tab:distances} illustrates that the ITDG resulting from floor reflections decreases as the distance between the sound source and the microphone increases. 
%The values in the table are achieved by setting the microphone height $h_s{=}1.5$m, matching the microphone height used in the STARSS23 dataset \cite{Shimada2023STARSS23AA}, and the source height $h_s{=}0.9$m, corresponding to the average event height in the STARSS23 training set, which aligns with the height of a seated user.
%We also include the case of sound events at 1.5\,m, which is a reasonable height for standing speakers. 
%While these conditions may not always apply, we argue that the model can learn prior knowledge about the typical height of sound sources based on their class. For instance, speech is unlikely to originate from the ceiling or floor, whereas footsteps are naturally associated with the ground.
%Ideally, the model should learn when and how to integrate such prior knowledge to enhance its distance estimation predictions.

For the second feature, we explore the role of early reflections in distance estimation, focusing on the ITDG, a key cue for perceiving distance \cite{Bronkhorst:1999:distPercep,kaplanis2014perception}. 
While early reflection delays also depend on room size, it is reasonable to assume that the earliest reflection originates from the floor \cite{kaplanis2014perception}, as shown in Fig.\,\ref{fig:floorReefl}.
%From this assumption, Table\,\ref{tab:distances} demonstrates that ITDG from floor reflections decreases as the source-microphone distance increases. These values assume a microphone height of $h_m{=}1.5$m, as in the STARSS23 dataset \cite{Shimada2023STARSS23AA}, and a source height of $h_s{=}0.9$m, reflecting the average event height in the training set, similar to a seated user. We also include sources at 1.5m, representing standing speakers.
Under this assumption, the ITDG from floor reflections decreases as the source–microphone distance grows. For instance, with a microphone height of $h_m{=}1.5$\,m (as in the STARSS23 dataset \cite{Shimada2023STARSS23AA}) and a source height of $h_s{=}0.9$\,m (average event height in the training set), the ITDG is 4.2\,ms at 1\,m, 3.0\,ms at 2\,m, and 2.3\,ms at 3\,m.
Although these conditions may not always hold, we argue that the model can learn prior knowledge about typical source heights based on class. For instance, speech is unlikely to originate from the ceiling or floor, whereas footsteps are naturally associated with the ground. Ideally, the model should determine when and how to incorporate such priors to refine distance estimation.

%------------------

\begin{figure}[tb]
\centerline{\includegraphics[width=\columnwidth]{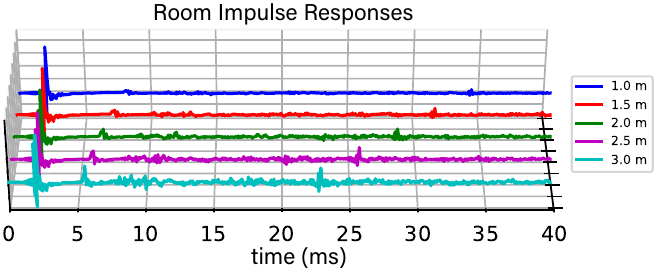}}
\caption{RIRs from the omnidirectional FOA channel of the SurrRoom 1.0 dataset \cite{cieciura:2023:SurrRoom} (``Pop\_Recording\_Studio'' room) used to spatialize speech at different distances. Direct sound peaks are temporally aligned for comparison.}
\label{fig:rirs}
\vspace{-3mm}
\end{figure}

To design an input feature that captures early reflections, we conducted a preliminary study on ITDG variations across different source distances. We analyzed an 8s speech clip from the S3A Object-based Audio Drama dataset \cite{cieciura:2024:turning, Woodstock:2016:changing} and convolved it with room impulse responses (RIRs) from SurrRoom 1.0 \cite{cieciura:2023:SurrRoom}. Specifically, we used the omnidirectional W channel of FOA RIRs recorded in the ``Pop\_Recording\_Studio'' at distances of [1m, 1.5m, 2m, 2.5m, 3m].
Fig.\,\ref{fig:rirs} shows the aligned RIRs, where the first reflection, i.e., the initial peak after the direct sound, shifts closer to the direct sound as distance increases. %consistent with Table\,\ref{tab:distances}.
A later strong reflection, likely from the rear wall, also appears, with increasing delay at greater distances. % (e.g., $\sim$23ms at 3m vs. $\sim$34ms at 1m).
While wall reflections depend on room size and geometry, making them unreliable for distance estimation, floor reflections offer a more robust and consistent cue for this task.

\begin{figure}[tb]
\centering
\begin{minipage}{\columnwidth}
% Answer: [trim={left bottom right top},clip]
\includegraphics[width=\columnwidth]{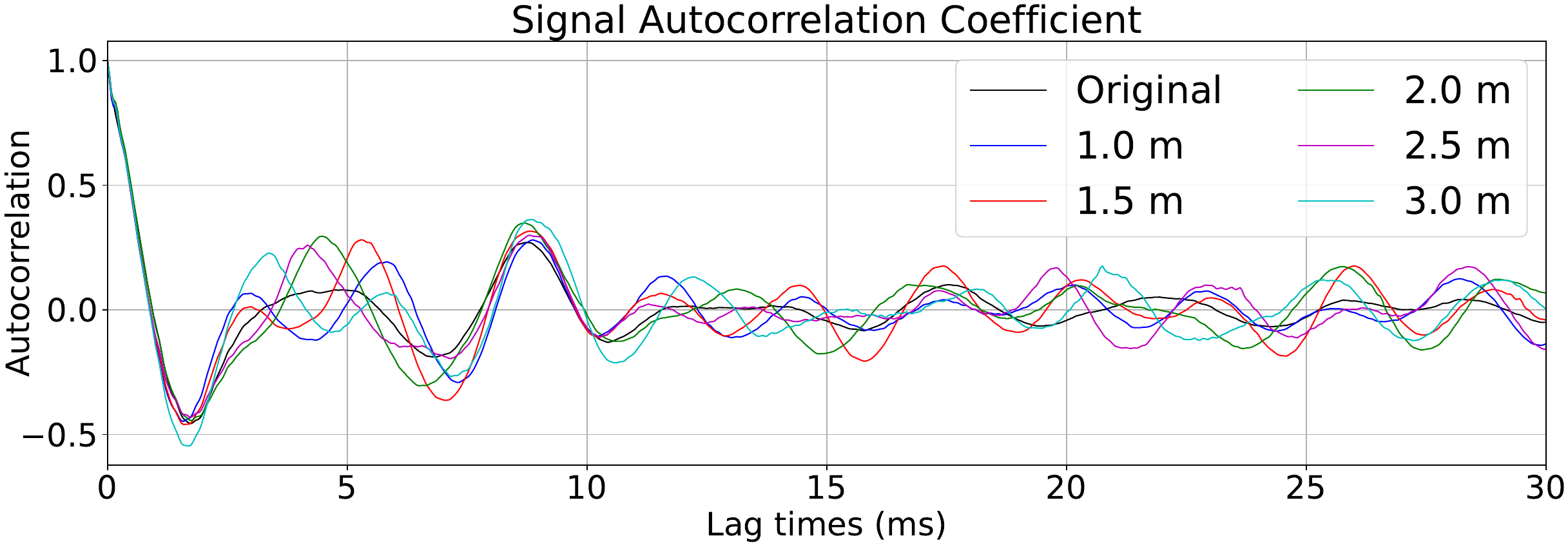}
%\vspace{-7mm}
%\captionof*{figure}{\footnotesize (a) ACC}
\end{minipage}\hfill

\begin{minipage}{\columnwidth}
\includegraphics[width=\columnwidth]{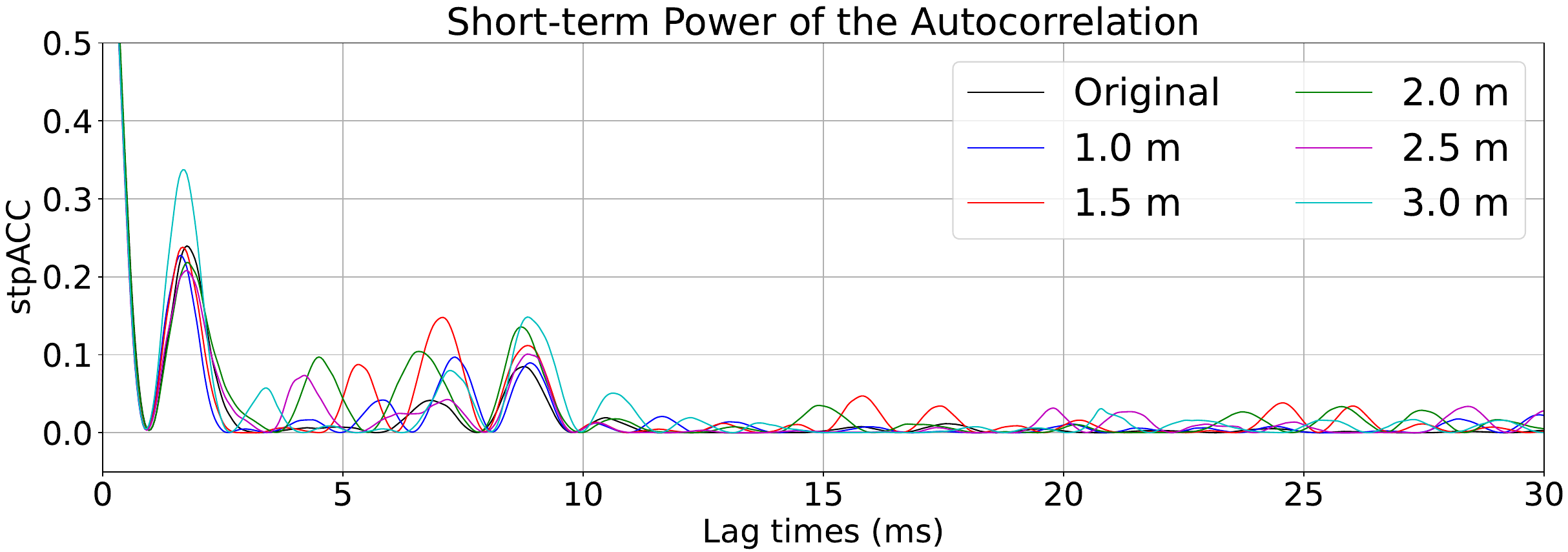} 
\vspace{-7mm}
%\captionof*{figure}{\footnotesize (b) STP}   
\end{minipage}
\vspace{3mm}
\captionof{figure}{Autocorrelation coefficient at varying distances (top). Short-term power of the autocorrelation (bottom). ``Original'' refers to the clean, unconvolved speech clip.}
\label{fig:acc_features} \vspace{-4mm}
\end{figure}%

%---------------------

After spatializing the clip at varying distances, we compute the correlation coefficient and extract its energy envelope. The upper part of Fig.\,\ref{fig:acc_features} illustrates the first 30\,ms of normalized autocorrelation coefficients (ACCs) across different distances. As expected, the second ACC peak appears earlier as the distance increases. %aligns closely with the ideal ITDG values from Table\,\ref{tab:distances} for $h_s{=}1.5$\,m. 
To strengthen this representation capturing both the level and timing of the first reflection, we compute the short-term power of the ACC (stpACC).
The stpACC is obtained by applying a Hann-windowed moving average (size: 8 samples) to the squared ACC coefficients. At 24 kHz sampling rate, this $\sim$0.3 ms window groups reflections from objects or surfaces within 10\,cm of each other. The resulting stpACC features for spatialized speech signals are shown in the lower part of Fig.\,\ref{fig:acc_features}.

To leverage stpACC features for the 3D SELD task and allow concatenation with conventional SELD features (e.g., IVs and log-mel spectrograms), we represent them as 2D signals. This is achieved by computing the short-time autocorrelation function in the frequency domain as:
\begin{equation}
    ACC(t,\tau)=\mathcal{F}^{-1}_{f\rightarrow\tau} (\mathbf{X}(t,f)\mathbf{X}^*(t,f))
\end{equation}
\begin{equation}
    ACC^{\mathrm{norm}}(t,\tau)= \dfrac{ACC(t,\tau)}{\mathrm{max}_\tau(|ACC(t,\tau)|)}, \hspace{3mm}\forall t
\end{equation}
%\begin{equation}
%    ACC = \dfrac{ACC}{\mathrm{max}(|ACC|)}
%\end{equation}
where $\mathbf{X}(t,f)$ is the STFT of the W channel, $(.)^*$ denotes complex conjugate, and $\mathcal{F}^{-1}_{f\rightarrow\tau}$ the inverse FFT from the frequency $f$ to the time-lag domain $\tau$.
We then normalize each time bin $t$ %by $\mathrm{max}_\tau(|ACC(t,\tau)|)$, 
so that $ACC^{\mathrm{norm}}(t,0){=}1$, and convolve its square with an 8-sample Hann window to obtain $stpACC(t,\tau)$.

\begin{table*}[tb]
\caption{Results with 95\% confidence intervals. All experiments are conducted using the CNN-Conformer architecture, while the fourth row is achieved with the ResNet-Conformer (RC). Best results are shown in bold; second-best are underlined.}
\begin{center}%\footnotesize
\begin{tabular}{c|c|c|c|c|c|c}
\hline
\textbf{\#} & \textbf{SED \& DOA Feat.}&\textbf{Distance Feat.}&$F_{\leq 20^{\circ}/1}\uparrow$& $DOAE\downarrow$& $RDE \downarrow$&\textit{SELD} $\downarrow$\\
\hline
1 & {SALSA-Lite} & {stpACC} & 29.1\% (25.0\%\,-\,32.7\%) & 22.4$^{\circ}$ (17.8$^{\circ}$\,-\,27.0$^{\circ}$) & 0.287 (0.248\,-\,0.321) & 0.373 (0.347\,-\,0.400) \\
\hline
2 & {SALSA} & {CDPD} \cite{Yeow:2025:CDPDdistFeat} & 28.3\% (24.1\%\,-\,32.3\%) & 24.3$^{\circ}$ (19.2$^{\circ}$\,-\,28.6$^{\circ}$) & 0.318 (0.286\,-\,0.352) & 0.390 (0.365\,-\,0.414) \\
3 & {SALSA-Lite} & {CDPD} \cite{Yeow:2025:CDPDdistFeat} & 27.6\% (23.5\%\,-\,31.1\%) & 22.5$^{\circ}$ (16.8$^{\circ}$\,-\,29.9$^{\circ}$) & 0.274 (0.243\,-\,0.301) & 0.374 (0.354\,-\,0.399) \\
\hline
4 & (RC) Log-Mels, IVs & {stpACC} & 29.6\% (25.1\%\,-\,33.4\%) & 22.7$^{\circ}$ (19.6$^{\circ}$\,-\,26.0$^{\circ}$) & 0.291 (0.265\,-\,0.316) & 0.374 (0.352\,-\,0.398) \\
\hline
5 & Log-Mels, IVs & - & 34.7\% (29.7\%\,-\,39.3\%) & \textbf{19.4$^{\circ}$} (16.6$^{\circ}$\,-\,22.2$^{\circ}$) & 0.296 (0.273\,-\,0.355) & 0.352 (0.332\,-\,0.385) \\
%& (29.7 - 39.3) & (16.6 - 22.2) & (0.273 - 0.355) & (0.332 - 0.385) \\
%\hline
6 & Log-Mels, IVs & {D+R} & \textbf{36.4\%} (31.1\%\,-\,41.6\%) & 22.0$^{\circ}$ (18.6$^{\circ}$\,-\,24.2$^{\circ}$) & 0.273 (0.234\,-\,0.298) & \underline{0.344} (0.314\,-\,0.367)\\
%\hline
7 & Log-Mels, IVs & {DRR} & \underline{36.0\%} (30.8\%\,-\,41.4\%) & \underline{20.1$^{\circ}$} (17.7$^{\circ}$\,-\,23.4$^{\circ}$) & 0.286 (0.240\,-\,0.315) & 0.346 (0.319\,-\,0.368)\\
%& (30.8 - 41.4) & (17.7 - 23.4) & (0.240 - 0.315) & (0.319 - 0.368)\\
%\hline
8 & Log-Mels, IVs & {stpACC} & 35.9\% (30.5\%\,-\,41.2\%) & 21.3$^{\circ}$ (11.9$^{\circ}$\,-\,30.6$^{\circ}$) & \underline{0.262} (0.225\,-\,0.296) & \textbf{0.341} (0.304\,-\,0.375) \\
%& (30.5 - 41.2) & (11.9 - 30.6) & (0.225 - 0.296) & (0.304 - 0.375) \\
%& (30.5 - 41.2) & (11.9 - 30.6) & (0.225 - 0.296) & (0.304 - 0.375) \\
\hline
9 & Log-Mels, IVs & {stpACC}, {D+R} & 33.0\% (28.4\%\,-\,37.2\%) & 22.6$^{\circ}$ (18.4$^{\circ}$\,-\,27.1$^{\circ}$) & 0.264 (0.227\,-\,0.294) & 0.353 (0.327\,-\,0.370) \\
10 & Log-Mels, IVs & {stpACC}, {DRR} & 31.7\% (27.1\%\,-\,35.3\%) & 24.0$^{\circ}$ (19.9$^{\circ}$\,-\,27.0$^{\circ}$) & \textbf{0.251} (0.217\,-\,0.277) & 0.356 (0.331\,-\,0.380) \\
\hline

\end{tabular}
\label{tab:results}
\end{center}
\vspace{-5mm}
\end{table*}

\section{Experiments}
\label{sec:experiments}

\subsection{Model Architecture}
\label{sec:model}

We evaluated the proposed distance features using a CNN-Conformer architecture, widely adopted for SELD \cite{Berghi:2024:ICASSP24,Wang:2023:ACS,Xue:2023:resnetConf,Berghi:2025:dcaseworkshop,Berghi:2025:DCASE25techRep}. It consists of a CNN encoder, a Conformer module \cite{Gulati2020ConformerCT}, and feed-forward layers for 3D SELD predictions. 
The CNN encoder processes acoustic features, including IVs in log-mel domain (3 channels), log-mel spectrograms from FOA (4 channels), and the proposed distance features, DRR, D+R, or stpACC, forming an input of shape $C_\mathrm{in}{\times} T_\mathrm{in}{\times} F_\mathrm{in}$. Here, $T_\mathrm{in}$ and $F_\mathrm{in}$ represent temporal and frequency (or time-lag) bins, respectively, with $C_\mathrm{in}{=}8$ for DRR and sptACC or $C_\mathrm{in}{=}9$ for D+R.
The CNN encoder comprises four convolutional blocks with residual connections, each containing two 3$\times$3 convolutional layers, BN, ReLU activation, and Avg pooling with a stride of 2, halving the temporal and frequency dimension at each block. The resulting tensor of shape $512\times T_\mathrm{in}/16\times F_\mathrm{in}/16$ is reshaped and frequency Avg pooling is applied to achieve a $T_\mathrm{in}/16\times 512$ embedding. 
$T_\mathrm{in}$ is chosen so that $T_\mathrm{in}/16$ matches the label frame rate (10 labels\,/\,sec).
The Conformer module includes eight layers and eight attention heads. Finally, two feedforward layers predict multi-ACCDDOA vectors, modeling $N{=}3$ tracks \cite{Krause:2024:seldDistance}. 
As in \cite{Diaz-Guerra:2024:seldBaseline24,Krause:2024:seldDistance}, the model is trained using class-wise Auxiliary Duplicating Permutation Invariant Training (ADPIT) loss \cite{Shimada:2022:multiACCDOA}.

\subsection{Dataset and Data Augmentation}
\label{sec:dataset}

Experiments are conducted on the STARSS23 dataset \cite{Shimada2023STARSS23AA}, %, which, to our knowledge, is the only public dataset for 3D SELD. Other well-known benchmarks for SELD, such as STARSS22 \cite{Politis2022DCASE} or TAU-NIGENS Spatial Sound Events 2020 and 2021 \cite{politis:2020:DCASE,politis:2021:DCASE}, do not include distance labels.
%STARSS23 \cite{Shimada2023STARSS23AA} consists of $\sim$7.5h of real spatial recordings of acoustic scenes, temporally and spatially annotated, with 13 event classes. 
%In our experiments, we employed the FOA audio format. 
%Audio in STARSS23 is captured with a 32-channel Eigenmike and is provided in MIC and FOA spatial formats.
which provides 3D SELD labels at 100ms resolution. 
%The dataset includes directional interferes, i.e., non-target sounds that should not be detected. 
%The dataset presents a development set and an evaluation set.
The development set, with its predefined train–test split, was used for evaluation. To expand training data and reduce overfitting, we applied audio channel swap (ACS) augmentation \cite{Wang:2023:ACS}, increasing the dataset eightfold.
Models were pre-trained on 20\,h of synthetic data from the DCASE2024 Task 3 Challenge \cite{Roman:2024:spatialScaper}, also with ACS augmentation, and then fine-tuned on real recordings. This strategy mitigates dataset imbalance, which could bias models toward synthetic conditions. As noted by Yeow \etal \cite{Yeow:2025:CDPDdistFeat}, synthetic data often improves F1 and localization but harms distance estimation due to mismatched distributions and limited real-world reverberation characteristics. Pre-training followed by fine-tuning alleviated these issues and consistently improved distance estimation.
%We pre-trained on synthetic data and fine-tuned on the real one rather than training from scratch on both due to the imbalance between datasets, which could bias the model toward the synthetic domain. As noted by Yeow \etal \cite{Yeow:2025:CDPDdistFeat}, synthetic data can boost F1 and localization but often harms distance estimation because of mismatched distance distributions and a lack of real-world reverberation characteristics. Pre-training followed by fine-tuning mitigated this issue and consistently improved distance estimation. 
Moreover, pre-training proved essential for analyzing the effect of our input features, likely because it provides prior knowledge of sound events, enabling better interpretation of feature information.

\begin{figure}[tb]
\centerline{\includegraphics[width=\columnwidth]{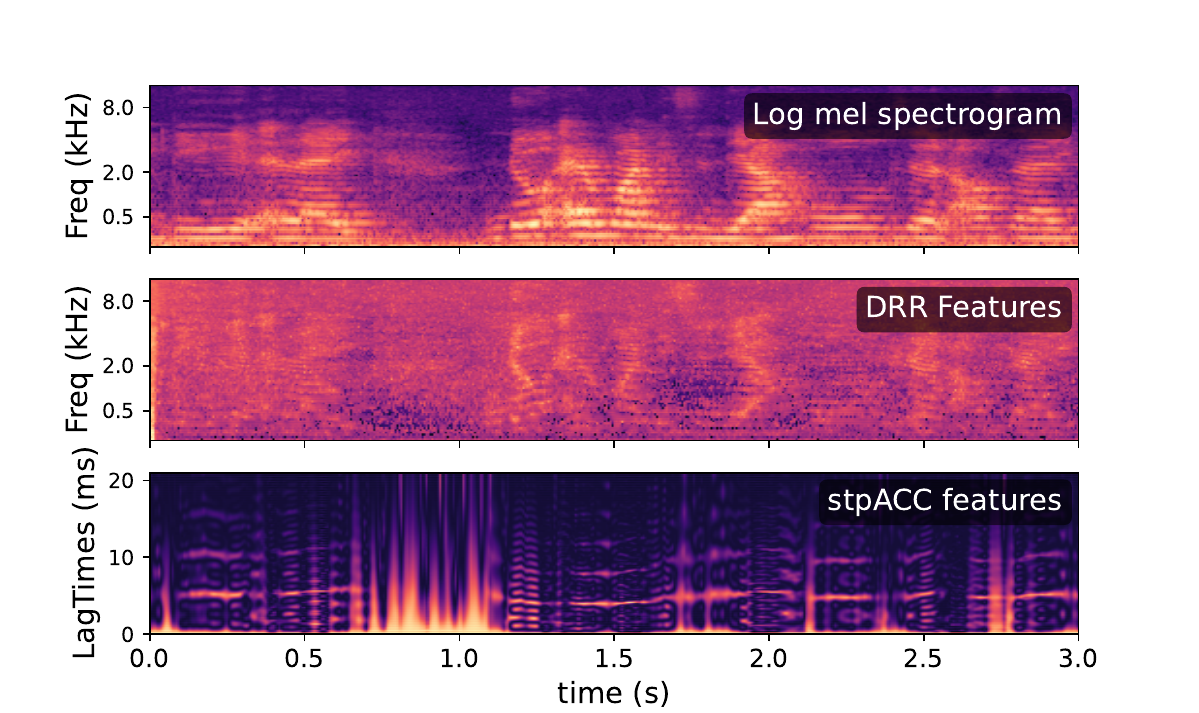}}
\caption{Distance features with respective log-mel spectrogram extracted from a sequence of STARSS23.}
\label{fig:featSpec}
\vspace{-4mm}
\end{figure}

\iffalse
\subsection{Metrics}
\label{sec:metrics}

To evaluate our models, we adopted the official metrics of the DCASE 2024 Task 3 Challenge \cite{Diaz-Guerra:2024:seldBaseline24} that are based on true positive (TP) and false positive (FP) predictions. A prediction is considered TP if the class prediction is correct and if its predicted DOA is within $\pm$20$^{\circ}$ from the target, and the relative distance error ($RDE{=}|L_p-L_r|/L_r$ \hl{is dimensionless}, with $L_p$ and $L_r$ being the predicted and reference distance, respectively) is smaller than 1. 
Metrics are computed at the frame level and for each class independently and then averaged across the number of classes.
Based on these, the adopted metrics are the class- and location-dependent F1 score ($F_{\leq 20^{\circ}/1}$), the class-dependent DOA error ($DOAE$), and the class-dependent relative distance error ($RDE$) \cite{Diaz-Guerra:2024:seldBaseline24}. We also include the $SELD$ score that encodes the overall 3D SELD performance and is achieved as: $SELD{=}\mathrm{mean}((1-F_{\leq 20^{\circ}/1}), DOAE/180, RDE)$.
%\begin{equation}
%    SELD=\mathrm{mean}((1-F_{\leq 20^{\circ}/1}), DOAE/180, RDE)
%\end{equation}
\fi

\subsection{Hyper-parameters and Experimental Settings}
\label{sec:exp_settings}

%To train our models, we divided the dataset into chunks of 3 seconds, extracted at steps of 1\,s for training and with no overlap for testing.
%An STFT with 512-point Hann window and hop size of 150 samples was used to generate spectrograms, discretizing the 3-second audio segments (24kHz) into 480 temporal bins ($T_\mathrm{in}$). We employed 128 frequency bins to compute the log-mel spectrograms for the audio channels, as well as for the DRR features, D+R features, and IVs. % in log-mel domain. 
%For the stpACC features, we applied an STFT with a 1014-point Hann window. This ensures that, when considering only the positive time-lags $\tau\,{>}\,0$, $stpACC(t,\tau)$ contains 512 time-lag bins, covering delays up to approximately 20\,ms after the direct sound. We then downsample the time-lag dimension by a factor of 4 to achieve 128 bins and allow concatenation with the other features.   
%We trained our models with batches of 32 inputs and Adam optimizer for 50 epochs, then we selected the best epoch based on the lowest SELD score. The learning rate is set to 5e-5 for the first 30 epochs, then and it is decreased by 5\% every epoch.

We trained models on 3‑second audio chunks, extracted every 1\,s for training and without overlap for testing. Spectrograms were computed via STFT (512‑point Hann window, 150‑sample hop), producing 480 time frames for 24\,kHz audio. Log‑mel spectrograms (128 bins) were generated for audio channels, DRR, D+R, and IV features. 
For stpACC features, an STFT with a 1014-point Hann window yielded 512 time‑lag bins.
This ensures that, when considering only the positive time-lags $\tau\,{>}\,0$, $stpACC(t,\tau)$ covers delays up to approximately 20\,ms after the direct sound. We then downsample the time-lag dimension by a factor of 4 to achieve 128 bins and allow concatenation with the other features.
%For stpACC, an STFT with a 1014-point Hann window produced 512 time-lag bins (up to ~20ms), later downsampled to 128 for feature concatenation. 
%Models were trained with batch size 32 using Adam optimizer for 50 epochs, selecting the best based on the lowest SELD score. The learning rate was 5e-5 for 30 epochs, then reduced by 5\% per epoch.
Models were trained for 50 epochs with Adam (batch size 32), selecting the best based on SELD score. The learning rate started at 5e‑5 for 30 epochs, then decayed by 5\% per epoch.

To evaluate our models, we adopted the official metrics of the DCASE 2024 Task 3 Challenge \cite{Diaz-Guerra:2024:seldBaseline24} %that are based on true positive (TP) and false positive (FP) predictions. A prediction is considered TP if the class prediction is correct and if its predicted DOA is within $\pm$20$^{\circ}$ from the target, and the relative distance error ($RDE{=}|L_p-L_r|/L_r$ \hl{is dimensionless}, with $L_p$ and $L_r$ being the predicted and reference distance, respectively) is smaller than 1.  Metrics are computed at the frame level and for each class independently and then averaged across the number of classes.
%Based on these, the adopted metrics are the class- and location-dependent F1 score ($F_{\leq 20^{\circ}/1}$), the class-dependent DOA error ($DOAE$),
They are the class- and location-dependent F1 score ($F_{\leq 20^{\circ}/1}$), the class-dependent DOA error ($DOAE$), and the class-dependent relative distance error ($RDE$) \cite{Diaz-Guerra:2024:seldBaseline24}. 
$RDE$ is dimensionless as it is defined as $RDE{=}|L_p-L_r|/L_r$, with $L_p$ and $L_r$ being the predicted and reference distance, respectively.
We also include the $SELD$ score that encodes the overall 3D SELD performance and is achieved as: $SELD{=}\mathrm{mean}((1-F_{\leq 20^{\circ}/1}), DOAE/180, RDE)$.

\subsection{Results}
\label{sec:results}

Table\,\ref{tab:results} summarizes the results obtained using the proposed distance features, with 95\% confidence intervals estimated via the jackknife variance method \cite{efron:1981:jackknife} through leave-one-out resampling across the 78 test sequences.
Beyond the baseline configuration, i.e., a CNN-Conformer trained on log-mel spectrograms, IVs, and one of our proposed features (rows 5–8), we conducted additional experiments. These include: combining autocorrelation and DRR-based features (rows 9–10); replacing FOA-derived log-mel spectrograms and IVs with MIC-based SALSA-Lite features \cite{Nguyen:2021:SALSALiteAF} (row 1); and evaluating CDPD distance features \cite{Yeow:2025:CDPDdistFeat}, designed for SALSA and SALSA-Lite (rows 2–3).
Additionally, given the popularity of ResNet-Conformer architectures in SELD research \cite{Wang:2023:ACS,dong:2025:anExperimStudy,Abolfazli:2024:3dseld_withresnet,Xue:2023:resnetConf}, row 4 reports results using stpACC features with a ResNet-Conformer (RC) instead of the CNN-Conformer.

Replacing log-mel and IVs with SALSA-Lite (row 1) degraded performance, likely due to spatial format differences: SALSA-Lite derives from MIC, whereas IVs originate from FOA, which typically performs better on DCASE datasets. We believe this is because FOA is analytically derived from a 32-channel Eigenmike, while MIC uses only a four-channel subset. This disparity may also explain the weaker performance of CDPD features. In contrast, our features work with both FOA and MIC, providing a clear advantage over CDPD.
Using a ResNet-Conformer instead of a CNN-Conformer (row 4) also reduced performance, possibly because the deeper architecture requires more extensive pre-training.

Results with log-mel spectrograms and IVs indicate that all proposed distance features contributed to reducing $RDE$, leading to overall $SELD$ score improvement. A small improvement was also observed in $F_{\leq 20^{\circ}/1}$. The model without distance features achieved the lowest $DOAE$. Notably, the $DOAE$ confidence interval with stpACC features is over three times larger than with other features, indicating higher variability and noisier estimates.
Despite this, stpACC features yielded the best $RDE$ among the proposed features and the highest $SELD$ score. While DRR is a known indicator of distance perception \cite{Griesinger:2009:DRR,chitreddy:2020:distPerc,Zahorik2005AuditoryDP,Kolarik:2016:AuditoryDP}, D+R features produced better distance estimates, possibly because the model benefits from greater flexibility in learning task-relevant information from the direct and reverberant components separately.
Combining stpACC with DRR further reduced $RDE$ to 0.251, but at the cost of lower F1 and localization scores, suggesting that stpACC alone offers the best overall 3D SELD balance.

\section{Conclusion}
\label{sec:conclusion}

This paper proposes novel acoustic features for distance estimation in 3D SELD using reverberation cues. The first feature category separates direct and reverberant components, either fed independently or as a direct-to-reverberant ratio. The second focuses on early reflections, particularly the first reflection.
Experiments on STARSS23 show that these features improve distance estimation and overall SELD score, with short-term autocorrelation power achieving state-of-the-art distance performance. Experiments included various combinations of FOA- and MIC-generated input features and model architectures.
%Future research will explore the effectiveness of reverberation-based distance features across different 3D SELD data, including synthetic data, and various network architectures.

\bibliographystyle{IEEEtran}
\bibliography{SPL2025_}

@inproceedings{politis:2021:DCASE,
    author = "{Politis {et al.}}, Archontis",
    title = "A Dataset of Dynamic Reverberant Sound Scenes with Directional Interferers for Sound Event Localization and Detection",
    booktitle = "DCASE Workshop",
    year = "2021",
    pages_ = "125--129",
    isbn = "978-84-09-36072-7",
    doi. = "10.5281/zenodo.5770113",
}

@inproceedings{politis:2020:DCASE,
  title = "A Dataset of Reverberant Spatial Sound Scenes with Moving Sources for Sound Event Localization and Detection",
  author = "Archontis {Politis et al.}",
  year = "2020",
  booktitle = "DCASE Workshop",
}

@ARTICLE{Wang:2023:ACS,
  author={{Wang {et al.}}, Qing},
  journal={IEEE/ACM TASLP}, 
  title={A Four-Stage Data Augmentation Approach to {ResNet-Conformer} Based Acoustic Modeling for Sound Event Localization and Detection}, 
  year={2023},
  volume={31},
  number={},
  pages={1251-1264},
  doi={10.1109/TASLP.2023.3256088}
}

@INPROCEEDINGS{dong:2025:anExperimStudy,
  author={Yuxuan {Dong et al.}},
  author_={Dong, Yuxuan and Wang, Qing and Hong, Hengyi and Jiang, Ya and Cheng, Shi},
  booktitle={ICASSP}, 
  booktitle_={ICASSP 2025 - 2025 IEEE International Conference on Acoustics, Speech and Signal Processing (ICASSP)}, 
  title={An Experimental Study on Joint Modeling for Sound Event Localization and Detection with Source Distance Estimation}, 
  year={2025},
  volume={},
}

@INPROCEEDINGS{cao:2021:EINv2,
  author_={Cao, Yin and Iqbal, Turab and Kong, Qiuqiang and An, Fengyan and Wang, Wenwu and Plumbley, Mark D.},
  author={{Cao {et al.}}, Yin},
  booktitle_={IEEE International Conference on Acoustics, Speech and Signal Processing}, 
  booktitle={ICASSP},   
  title={An Improved Event-Independent Network for Polyphonic Sound Event Localization and Detection}, 
  year={2021},
  volume={},
  number={},
  pages_={885-889},
  doi={10.1109/ICASSP39728.2021.9413473}
}

@mastersthesis{Sheeline:1982:DRR,
	title = {An Investigation of the Effects of Direct and Reverberant Signal Interactions on Auditory Distance Perception},
	volume = {Ph.D.},
	number = {STAN-M-13},
	year = {1982},
	school = {Stanford University},
	type = {Ph.D.},
	url = {https://ccrma.stanford.edu/files/papers/stanm13.pdf},
	author = {Christopher Sheeline}
}

@INPROCEEDINGS{Berghi:2023:WASPAA,
  author = {Berghi, Davide and Jackson, Philip J. B.},
  title={Audio Inputs for Active Speaker Detection and Localization via Microphone Array}, 
  year={2023},
  booktitle={IEEE Workshop on Applications of Signal Processing to Audio and Acoustics},
}

@article{Bronkhorst:1999:distPercep,
    author = {Adelbert W. Bronkhorst and Tammo Houtgast},
    title = {Auditory distance perception in rooms},
    journal = {Nature},
    volume = {397},
    pages = {517-520},
    year = {1999}
}

@article{Zahorik2005AuditoryDP,
  title={Auditory distance perception in humans: A summary of past and present research},
  author={P. Zahorik and D. Brungart and A. Bronkhorst},
  journal={Acta Acustica United With Acustica},
  year={2005},
  volume={91},
  pages={409-420}
}

@article{Kolarik:2016:AuditoryDP,
    author = {Kolarik, A.J. and Moore, B.C.J. and Zahorik, P. and Cirstea, S. and Pardhan, S.},
    title = {Auditory distance perception in humans: a review of cues, development, neuronal bases, and effects of sensory loss},
    journal = {Atten Percept Psychophys},
    volume = {78},
    pages={373–395},
    year = {2016}
}

@inproceedings{cieciura:2024:turning,
    author = {Craig Cieciura and Elettra Bargiacchi and P. J. B. Jackson},
    title = {Authoring inter-compatible flexible audio for mass personalization},
    booktitle = {The 157th Audio Engineering Society Convention},
    year = {2024}
}

@inproceedings{Diaz-Guerra:2024:seldBaseline24,
title = "Baseline models and evaluation of sound event localization and detection with distance estimation in {DCASE} 2024 {C}hallenge",
author = "{Diaz-Guerra {et al.}}, David",
year = "2024",
pages = "41--45",
booktitle = "DCASE Workshop",
}

@ARTICLE{Lu:2010:binauralDist,
  author={Lu, Yan-Chen and Cooke, Martin},
  journal={IEEE Transactions on Audio, Speech, and Language Processing}, 
  title={Binaural Estimation of Sound Source Distance via the Direct-to-Reverberant Energy Ratio for Static and Moving Sources}, 
  year={2010},
  volume={18},
  number={7},
  pages={1793-1805}
}

@inproceedings{Gulati2020ConformerCT,
  title={Conformer: Convolution-augmented Transformer for Speech Recognition},
  author={Anmol {Gulati {et al.}}},
  year={2020},
  booktitle={Interspeech},
  pages={5036--5040}
}

@article{Roman:2024:ehnancedAVseld,
  title={Enhanced Sound Event Localization and Detection in Real 360-degree audio-visual soundscapes},
  author_={Adrian S. Roman and Baladithya Balamurugan and Rithik Pothuganti},
  author={Adrian S. {Roman {et al.}}},
  journal={ArXiv},
  year={2024},
  volume={abs/2401.17129}
}

@ARTICLE{Yeow:2025:CDPDdistFeat,
  author_={Yeow, Jun-Wei and Tan, Ee-Leng and Bai, Jisheng and Peksi, Santi and Gan, Woon-Seng},
  author={{Yeow {et al.}}, Jun-Wei},
  journal={IEEE Sensors Journal}, 
  title={Enhancing 3-D Sound Event Localization and Detection With Distance Estimation Using Reverberation and Spatial Coherence Features}, 
  year={2025},
  volume={25},
  number={15},
  pages={29221-29237},
  doi={10.1109/JSEN.2025.3583033}}

@InProceedings{cao:2020:EIN,
  title={Event-Independent Network for Polyphonic Sound Event Localization and Detection},
  author_={Cao, Yin and Iqbal, Turab and Kong, Qiuqiang and Zhong, Yue and Wang, Wenwu and Plumbley, Mark D},
  author={Cao {et al.}, Yin},
  booktitle_={Detection and Classification of Acoustic Scenes and Events Workshop},
  booktitle={DCASE Workshop},
  year={2020}
}

@INPROCEEDINGS{Jiang:2024:AVseld,
  author={{Jiang {et al.}}, Ya},
  booktitle={ICME}, 
  title={Exploring Audio-Visual Information Fusion for Sound Event Localization and Detection In Low-Resource Realistic Scenarios}, 
  year={2024},
  volume={},
  number={},
  pages={1-6},
  doi={10.1109/ICME57554.2024.10687782}
}

@inproceedings{Berghi:2024:ICASSP24,
    author = {{Berghi {et al.}}, Davide},
    title = {Fusion of audio and visual embeddings for sound event localization and detection},
    booktitle = {ICASSP},
    year = {2024},
}

@ARTICLE{Takuya:2012:WPE,
  author={Yoshioka, Takuya and Nakatani, Tomohiro},
  journal={IEEE Transactions on Audio, Speech, and Language Processing}, 
  title={Generalization of Multi-Channel Linear Prediction Methods for Blind MIMO Impulse Response Shortening}, 
  year={2012},
  volume={20},
  number={10},
  pages={2707-2720},
  doi={10.1109/TASL.2012.2210879}
}

@INPROCEEDINGS{Abolfazli:2024:3dseld_withresnet,
  author={Abolfazli, Zahra and Abutalebi, Hamid Reza and Virtanen, Tuomas},
  author_={Zahra {Abolfazli et al.}},
  booktitle_={2024 10th International Conference on Signal Processing and Intelligent Systems (ICSPIS)},
  booktitle={ICSPIS},
  title={Improving Distance Estimation in Sound Event Localization and Detection Using {ResNet50} and Multi-{ACCDDOA}}, 
  year={2024},
  volume={},
  number={},
}

@inproceedings{Berghi:2025:dcaseworkshop,
    author = "Berghi, Davide and Jackson, Philip J. B.",
    title = "Integrating Spatial and Semantic Embeddings for Stereo Sound Event Localization in Videos",
    booktitle = "DCASE Workshop",
    year = "2025",
}

@ARTICLE{Mershon:1975:revDist,
  author={Mershon, Donald H. and King, L. Edward},
  journal={Perception \& Psychophysics}, 
  title={Intensity and reverberation as factors in the auditory perception of egocentric distance}, 
  year={1975},
  volume={18},
  number={},
  pages={409-415},
  doi={10.3758/BF03204113}
}

@inproceedings{Berghi:2024:DCASE24techRep,
      title={Leveraging Reverberation and Visual Depth Cues for Sound Event Localization and Detection with Distance Estimation}, 
      author={Davide Berghi and Philip J. B. Jackson},
      year={2024},
      booktitle={Techical Report of DCASE Challenge}, 
}

@INPROCEEDINGS{Shimada:2022:multiACCDOA,
  author={{Shimada {et al.}}, Kazuki},
  booktitle={ICASSP}, 
  title={Multi-{ACCDOA}: {L}ocalizing And Detecting Overlapping Sounds From The Same Class With Auxiliary Duplicating Permutation Invariant Training}, 
  year={2022},
  volume={},
  number={},
  pages={316-320},
  doi={10.1109/ICASSP43922.2022.9746384}
}

@article{hong:2024:MVAnet,
  title={{MVANet}: {M}ulti-Stage Video Attention Network for Sound Event Localization and Detection with Source Distance Estimation},
  author={Hengyi {Hong {et al.}}},
  journal={ArXiv},
  year={2024},
  volume={abs/2411.14153}
}

@INPROCEEDINGS{Drude:2018:nara_wpe,
  author={Drude, Lukas and Heymann, Jahn and Boeddeker, Christoph and Haeb-Umbach, Reinhold},
  booktitle={Speech Communication; 13th ITG-Symposium}, 
  title={{NARA-WPE}: {A} {P}ython package for weighted prediction error dereverberation in {N}umpy and {T}ensorflow for online and offline processing}, 
  year={2018},
  volume={},
  number={},
  pages={1-5},
  doi={}
}

@inproceedings{kaplanis2014perception,
  author={Neofytos Kaplanis and Søren Bech and Søren Jensen Holdt and Toon van Waterschoot},
  booktitle={Audio Engineering Society Conference},
  title={Perception of reverberation in small rooms: {A} literature study},
  year={2014},
}

@inproceedings{Woodstock:2016:changing,
    author = {James Woodcock and Chris Pike and Frank Melchior and Philip Coleman and Andreas Franck and Adrian Hilton},
    title = {Presenting the {S3A} Object-Based Audio Drama Dataset},
    booktitle = {The 140th Audio Engineering Society Convention},
    year = {2016}
}

@INPROCEEDINGS{Xue:2023:resnetConf,
  author={{Xue {et al.}}, Lihua},
  booktitle={Int.\ Conf.\ on Wireless Comms. \& Sig.\ Proc.}, 
  title={{ResNet-Conformer} Network using Multi-Scale Channel Attention for Sound Event Localization and Detection in Real Scenes}, 
  year={2023},
  volume={},
  number={},
}

@ARTICLE{Nguyen:2021:SALSA,
  author={Nguyen, Thi Ngoc Tho and Watcharasupat, Karn N. and Nguyen, Ngoc Khanh and Jones, Douglas L. and Gan, Woon-Seng},
  journal={IEEE/ACM Trans. on Audio, Speech, and Language Processing}, 
  title={{SALSA}: {S}patial Cue-Augmented Log-Spectrogram Features for Polyphonic Sound Event Localization and Detection}, 
  year={2022},
  volume={30},
  number={},
  pages={1749-1762},
  doi={10.1109/TASLP.2022.3173054}
}

@INPROCEEDINGS{Nguyen:2021:SALSALiteAF,
  author_={Tho Nguyen, Thi Ngoc and Jones, Douglas L. and Watcharasupat, Karn N. and Phan, Huy and Gan, Woon-Seng},
  author={{Tho Nguyen {et al.}}, Thi Ngoc},
  booktitle_={IEEE International Conference on Acoustics, Speech and Signal Processing}, 
  booktitle={ICASSP}, 
  title={{SALSA}-{L}ite: {A} Fast and Effective Feature for Polyphonic Sound Event Localization and Detection with Microphone Arrays}, 
  year={2022},
  volume={},
  number={},
  pages={716-720},
  doi={10.1109/ICASSP43922.2022.9746132}
}

@INPROCEEDINGS{Krause:2024:seldDistance,
  author={{Krause {et al.}}, Daniel Aleksander},
  booktitle={EUSIPCO}, 
  title={Sound Event Detection and Localization with Distance Estimation}, 
  year={2024},
  volume={},
  number={},
  pages={286-290},
  doi={10.23919/EUSIPCO63174.2024.10715220}
}

@article{Adavanne:2019:SELDnet,
  title={Sound Event Localization and Detection of Overlapping Sources Using Convolutional Recurrent Neural Networks},
  author={Sharath {Adavanne et al.}},
  journal={IEEE J.\ of Selected Topics in Sig.\ Proc.},
  year={2019},
  volume={13}, 
  pages={34-48}
}

@INPROCEEDINGS{Sato:2025:physicsInformedDist,
  author_={Sato, Nao and Yasuda, Masahiro and Saito, Shoichiro and Harada, Noboru},
  author={Nao {Sato et al.}},
  booktitle_={ICASSP 2025 - 2025 IEEE International Conference on Acoustics, Speech and Signal Processing (ICASSP)}, 
  booktitle={ICASSP}, 
  title={Sound Source Distance Estimation Utilizing Physics-informed Prior for Sound Event Localization and Detection}, 
  year={2025},
  volume={},
  number={},
  pages={1-5},
}

@inproceedings{chitreddy:2020:distPerc,
  TITLE = {{Source Distance Perception with Reverberant Spatial Audio Object Reproduction of Real Rooms}},
  AUTHOR = {Chitreddy, Sandeep and Jackson, Philip},
  BOOKTITLE = {{Forum Acusticum}},
  PAGES = {2079-2086},
  YEAR = {2020},
}

@inproceedings{Berghi:2025:DCASE25techRep,
      title={Spatial and semantic embedding integration for stereo sound event localization and detection in regular videos}, 
      author={Davide Berghi and Philip J. B. Jackson},
      year={2025},
      booktitle={Techical Report of DCASE Challenge}, 
}

@INPROCEEDINGS{Roman:2024:spatialScaper,
  author={{Roman {et al.}}, Iran R.},
  booktitle={ICASSP}, 
  title={Spatial {S}caper: {A} Library to Simulate and Augment Soundscapes for Sound Event Localization and Detection in Realistic Rooms}, 
  year={2024},
  volume={},
  number={},
  doi={10.1109/ICASSP48485.2024.10446118}
}

@ARTICLE{Georganti:2011:distDet,
  author={Georganti, Eleftheria and May, Tobias and van de Par, Steven and Harma, Aki and Mourjopoulos, John},
  journal={IEEE Transactions on Audio, Speech, and Language Processing}, 
  title={Speaker Distance Detection Using a Single Microphone}, 
  year={2011},
  volume={19},
  number={7},
  pages={1949-1961},
}

@inproceedings{Shimada2023STARSS23AA,
  title={{STARSS}23: An Audio-Visual Dataset of Spatial Recordings of Real Scenes with Spatiotemporal Annotations of Sound Events},
  author={Kazuki {Shimada et al.}},
  booktitle={{NeurIPS}},
  year={2023},
}

@inproceedings{cieciura:2023:SurrRoom,
	author = {Cieciura, Craig and Volino, Marco and Jackson, Philip J. B.},
	booktitle = {{{The 154th Audio Engineering Society Convention}}},
	doi = {10:15126=surreydata:900689},
	langid = {english},
	title = {{{SurrRoom}} 1.0 {{Dataset}}: {S}patial Room Capture with 
	Controlled Acoustic and Optical Measurements},
	year = {2023},
}

@ARTICLE{Knapp:gccphat:1976,
  author={Knapp, Charles and Carter, G. Clifford},
  journal={IEEE Transactions on Acoustics, Speech, and Signal Processing}, 
  title={The generalized correlation method for estimation of time delay}, 
  year={1976},
  volume={24},
  number={4},
  pages={320-327},
  doi={10.1109/TASSP.1976.1162830}
}

@article{Griesinger:2009:DRR,
    author = {Griesinger, David},
    title = {The importance of the direct to reverberant ratio in the perception of distance, localization, clarity, and envelopment, Part one.},
    journal = {The Journal of the Acoustical Society of America},
    volume = {125},
    pages = {2483-2483},
    year = {2009},
    issn = {0001-4966},
    doi = {10.1121/1.4783276},
}

@article{efron:1981:jackknife,
 author = {B. Efron and C. Stein},
 journal = {The Annals of Statistics},
 number = {3},
 pages = {586--596},
 publisher = {Institute of Mathematical Statistics},
 title = {The Jackknife Estimate of Variance},
 urldate = {2025-02-06},
 volume = {9},
 year = {1981}
}

\end{document}